\documentclass[11pt]{article}
\usepackage{amssymb}
\usepackage{amsmath}
\usepackage{ifthen}

\catcode`@=11
\renewcommand{\section}{\@startsection{section}{1}{0pt}{\medskipamount}
{\medskipamount}{\large\bf}}
\numberwithin{equation}{section}
\catcode`@=12

\def\a{\alpha}

\def\de{\delta}
\def\eps{\epsilon}
\def\ve{\varepsilon}
\def\z{\zeta}

\def\l{\lambda}

\def\r{\rho}
\def\s{\sigma}

\def\ps{\psi}

\def\t{\tau}
\def\P{\Phi}
\def\L{\Lambda}
\def\G{\Gamma}

\newcommand{\C}{\mathbb C}
\newcommand{\R}{\mathbb R}
\newcommand{\Z}{\mathbb Z}

\def\Ne1{$N\,{=}\,1$}
\def\N2{$N\,{=}\,2$}
\def\Ng4{$N\,{=}\,4$}
\def\pa{\partial}

\def\tr{{\rm tr}}
\def\Kpmb#1{\setbox0=\hbox{${#1}$}   \kern-.025em\copy0\kern-\wd0
    \kern.05em\copy0\kern-\wd0     \kern-.025em\raise.0433em\box0 }

\newcommand{\ab}{{\bar{a}}}
\newcommand{\bb}{{\bar{b}}}
\newcommand{\cb}{{\bar{c}}}

\newcommand{\lb}{\bar{\l}}
\newcommand{\mb}{\bar{\mu}}
\newcommand{\nb}{{\bar{0}}}
\newcommand{\ob}{{\bar{1}}}
\newcommand{\ub}{\bar{u}}

\newcommand{\Zb}{\bar{Z}}

\newcommand{\bsy}[1]{\boldsymbol{#1}}
\newcommand{\ov}[1]{\overline{#1}}
\newcommand{\Psb}{\ov{\Psi}}
\newcommand{\Tb}{\ov{T}}

\newcommand{\zb}{\ov{z}}
\newcommand{\wb}{\ov{w}}
\newcommand{\at}{\widetilde{\a}}

\newcommand{\Gat}{\widetilde{\G}}
\newcommand{\Gt}{\widetilde{G}}
\newcommand{\lt}{\tilde{\l}}

\newcommand{\Phh}{{\widehat{\Phi}}}
\newcommand{\Psh}{\widehat{\Psi}}

\newcommand{\Gp}{\ifthenelse{\boolean{mmode}}{{G^+}}{\mbox{$G^+\:$}}}
\newcommand{\Gtp}{\ifthenelse{\boolean{mmode}}{\mbox{$\Gt^+$}}{\mbox{$\Gt^+\:$}}}
\newcommand{\Gm}{\ifthenelse{\boolean{mmode}}{{G^-}}{\mbox{$G^-\:$}}}
\newcommand{\Gtm}{\ifthenelse{\boolean{mmode}}{\mbox{$\Gt^-$}}{\mbox{$\Gt^-\:$}}}

\begin{document}
\begin{titlepage}
\setcounter{page}{0}
\begin{flushright}
hep-th/0204155\\
ITP--UH--07/02\\
YITP--SB--02--21\\
\end{flushright}

\vskip 2.0cm

\begin{center}

{\Large\bf Exact Solutions of Berkovits' String Field Theory}

\vspace{14mm}

{\large Olaf Lechtenfeld$\,^\circ$, Alexander D.\ Popov$\,^*$
	\ and \ Sebastian Uhlmann}
\\[5mm]
{\em Institut f\"ur Theoretische Physik  \\
Universit\"at Hannover \\
Appelstra\ss{}e 2, 30167 Hannover, Germany }\\[2mm]
{Email: lechtenf, popov, uhlmann@itp.uni-hannover.de}
\\[5mm]
{${}^\circ\ $
\em C.N. Yang Institute for Theoretical Physics \\
State University of New York \\
Stony Brook, NY 11794-3840, USA }

\end{center}

\vspace{2cm}

\begin{abstract}
\noindent
The equation of motion for Berkovits' WZW-like open (super)string field theory
is shown to be integrable in the sense that it can be written as the
compatibility condition (``zero-curvature condition'') of some linear
equations. Employing a generalization of solution-generating techniques
(the splitting and the dressing methods), we demonstrate how to construct
nonperturbative classical configurations of both \Ne1 superstring and \N2
fermionic string field theories. With and without $u(n)$ Chan-Paton factors,
various solutions of the string field equation are presented explicitly.
\end{abstract}

\vfill

\textwidth 6.5truein
\hrule width 5.cm
\vskip.1in

{\small
\noindent ${}^*$
On leave from Bogoliubov Laboratory of Theoretical Physics, JINR,
Dubna, Russia}

\end{titlepage}


\section{Introduction}
\noindent
String field theory opens new possibilities of exploring nonperturbative
structures in string theory. Starting from Witten's bosonic string field
theory action~\cite{Witten:1985cc} there have been several attempts to
construct a generalization to the supersymmetric case~\cite{Witten:1986qs,
Preitschopf:fc, Arefeva:1989cp} (for a general discussion of the subject
see~\cite{Preitschopf:fc, Thorn:1988hm}). The crucial novelty in the
superstring case is the appearance of the so-called picture degeneracy.
Each physical state is now represented by an infinite number of BRST-invariant
vertex operators in the covariant RNS~formalism~\cite{Friedan:1985ge}.
This enforces the choice of a definite picture number for each vertex
operator; for open superstring fields, one usually chooses all Ramond~(R)
string fields to carry $-\frac{1}{2}$ and all Neveu-Schwarz~(NS) string
fields to carry $-1$ units of picture charge.
As the total picture number must equal $-2$ for open superstrings and
Witten's bosonic Chern-Simons-like action contains a quadratic and a cubic
term in the string fields, this entails the introduction of picture
raising and lowering operators (carrying picture number $+1$ and $-1$,
respectively). The resulting action~\cite{Witten:1986qs}, however, is not
gauge-invariant~\cite{Wendt:1987zh, Lechtenfeld:1988tr} due to contact
term divergences occuring when two picture raising operators collide.
Although one can choose other pictures for the string fields~%
\cite{Preitschopf:fc,Arefeva:1989cp}, there is no choice for which the
action is cubic and gauge-invariant~\cite{Berkovits:1991gj}.

As an alternative, Berkovits proposed a (nonpolynomial) WZW-like
action~\cite{Berkovits:1995ab} for the NS sector which contains in its
Taylor expansion a kinetic term and a cubic interaction similar to
Witten's superstring field theory action:
\begin{equation}
  S=\frac{1}{2g^2}\,\tr\left\{ (e^{-\P}\Gp e^\P)(e^{-\P}\Gtp e^\P)
    -\int_0^1 dt (e^{-\Phh}\pa_t e^\Phh) \{ e^{-\Phh}\Gp e^\Phh,
    e^{-\Phh}\Gtp e^\Phh \} \right\} \, . \label{eq_BSSFTact}
\end{equation}
Here $e^\P={\cal I}+\P+\frac{1}{2}\P\star\P+\ldots$ is defined via Witten's
midpoint gluing prescription (${\cal I}$ denotes the identity string field),
$\P$ is a NS string field carrying $u(n)$ Chan-Paton labels\footnote{All
states in the string field $\P$ are taken to be GSO(+). For including
GSO($-$) states (e.\,g.\ for the study of tachyon condensation) one has
to add internal Chan-Paton labels to $\P$ and also to \Gp and \Gtp (for
a review on this subject see~\cite{Ohmori:2001am, DeSmet:2001af}).} with
an extension $\Phh(t)$ interpolating between $\Phh(t=0)=0$ and $\Phh(t=1)
=\P$. The BRST-like currents \Gp and \Gtp are the two superpartners of the
energy-momentum tensor in a twisted small superconformal \Ng4 algebra
with positive $U(1)$-charge. The action of \Gp and \Gtp on any string field
is defined in conformal field theory language as taking the contour integral,
e.\,g.\
\begin{equation}
  (\Gp e^\P)(z) = \oint \frac{dw}{2\pi i}\, \Gp(w) e^\P(z)\, , \quad
    (\Gtp e^\P)(z) = \oint \frac{dw}{2\pi i}\, \Gtp(w) e^\P(z)\, ,
    \label{eq_opact}
\end{equation}
with the integration contour running around $z$.

The action~(\ref{eq_BSSFTact}) is invariant under the transformation
\begin{equation}
  \delta e^\P = \L e^\P + e^\P \tilde{\L} \qquad\mbox{with}\qquad
    \Gp\L = 0 \, ,\quad \Gtp \tilde{\L}=0\, , \label{eq_BSSFTgau}
\end{equation}
and arguments based on this gauge invariance suggest that beyond reproducing
the correct four-point tree amplitude all $N$-point tree amplitudes are
correctly reproduced by~(\ref{eq_BSSFTact}). The corresponding equation
of motion reads
\begin{equation}
  \Gtp(e^{-\P} \Gp e^\P)=0 \, , \label{eq_BSSFTeom}
\end{equation}
where contour integrations are implied again.
This is the equation we set out to solve.

Solutions to the string field equation~(\ref{eq_BSSFTeom}) describe
classical string fields, which convey information about nonperturbative
string configurations. There has been renewed interest in such solutions
for open string field theories, initiated by a series of conjectures due
to Sen (see~\cite{Sen:1999mg} for a review and a list of references).
The latter have been tested in various ways, in particular within the above-%
described nonpolynomial superstring field theory: Using the level-truncation
scheme, numerical checks were performed e.\,g.\ in~\cite{Berkovits:2000zj,
Berkovits:2000hf, DeSmet:2000dp, Iqbal:2000st}, and the predicted kink
solutions describing lower-dimensional D-branes were 
found~\cite{Ohmori:2001sx}.
On the analytic side, a background-independent version of Berkovits' string
field theory was proposed in~\cite{Kluson:2001kk, Sakaguchi:2001kk},
and the ideas of vacuum string field theory and computations of the sliver
state of bosonic string field theory were transferred to the superstring
case~\cite{Marino:2001ny, Arefeva:2002mb, Ohmori:2002}. More recently, there
have been some attempts~\cite{Kluson:2001sb, Kluson:2002kk, Ohmori:2002} to
solve the string field equation~(\ref{eq_BSSFTeom}) but, to our knowledge,
no general method for finding explicit solutions has been presented so far.

In the present paper we show that the WZW-like string field equation~%
(\ref{eq_BSSFTeom}) is integrable in the sense that it can be written as the
compatibility condition of some linear equations with an extra ``spectral
parameter'' $\l$. This puts us in the position to parametrize solutions
of~(\ref{eq_BSSFTeom}) by solutions of linear equations on extended string
fields (depending on the parameter $\l$) and to construct classes of
explicit solutions via various solution-generating techniques. This
discussion will be independent of any implementation of \Gp and \Gtp in
terms of matter multiplets and is therefore valid for \Ne1 superstrings
as well as for \N2 strings.

We discuss two related approaches to generating solutions of the WZW-like
string field equation~(\ref{eq_BSSFTeom}). First, considering the
splitting method and using the simplest Atiyah-Ward ansatz for the matrix-%
valued string field of the associated Riemann-Hilbert problem, we reduce
eq.~(\ref{eq_BSSFTeom}) to the linear equations $\Gtp\Gp\r_k=0$ with $k=0,
\pm 1$. Here $\r_0$ and $\r_{\pm 1}$ are some string fields parametrizing
the field $e^\P$. However, our discussion of the splitting approach is
restricted to the $n=2$ case, i.\,e.\ $u(2)$ Chan-Paton factors, or a
certain embedding of $u(2)$ into $u(2+l)$. Second, we consider the (related)
dressing approach which overcomes this drawback. With this method, new
solutions are constructed from an old one by successive application of
simple (dressing) transformations. Using them, we write down an ansatz
which reduces the nonlinear equation of motion~(\ref{eq_BSSFTeom}) to a
system of linear equations. Solutions of the latter describe nonperturbative
field configurations obeying~(\ref{eq_BSSFTeom}). Finally, we present
some explicit solutions of the WZW-like string field equation.

The paper is organized as follows: In the next section we describe the
superconformal algebras needed for the formulation of the action~%
(\ref{eq_BSSFTact}) and give their implementation in terms of massless
\Ne1 and \N2 matter multiplets for later use. In section~\ref{sec_RealProp}
we discuss the reality properties of some operators in the superconformal
algebra. This issue is crucial for the treatment of the linear equations.
We prove the integrability of the WZW-like string field equation in section~%
\ref{sec_Integr}. The construction of solutions via solving a Riemann-Hilbert
problem and via dressing a seed solution is subject to sections~%
\ref{sec_SplitAppr} and~\ref{sec_DressAppr}, respectively.
Section~\ref{sec_sol} presents some explicit solutions.  Finally
section~\ref{sec_Concl} concludes with a brief summary and open problems.

\smallskip
\section{Definitions and conventions}
\noindent
{\bf Small $\bsy{N\,{=}\,4}$ superconformal algebra.} To begin with,
we consider an \N2 superconformal algebra generated by an energy-momentum
tensor $T$, two spin~3/2 superpartners $G^\pm$ and a $U(1)$ current%
\footnote{The labels denote the $U(1)$-charge w.\,r.\,t.\ $J$.} $J$. It
can be embedded into a small \Ng4 superconformal algebra with two
additional superpartners $\Gt^\pm$ and two spin~1 operators $J^{++}$ and
$J^{--}$ supplementing $J$ to an $SU(2)$ (or $SU(1,1)$) current algebra%
\footnote{The case of an $SU(2)$ current algebra applies to \Ne1 strings,
and the case of an $SU(1,1)$ current algebra corresponds to the \N2 string.}.
To this end~\cite{Ohmori:2001am}, $J$ can be ``bosonized'' as $J=\pa H$, and
we define 
\begin{equation}
  J^{++}:=e^H, \quad J^{--}:=e^{-H}\, ,
\end{equation}
where $H(z)$ has the OPE
\begin{equation}
  H(z) H(0) \sim \frac{c}{3} \log z
\end{equation}
with a central charge $c$. Then $\Gt^\pm$ can be defined by
\begin{equation}
  \Gt^-(z):=\oint \frac{dw}{2\pi i} J^{--}(w) \Gp(z) =
    [J_0^{--},\Gp(z)] \, , \label{eq_GtmDef}
\end{equation}
\begin{equation}
  \Gt^+(z):=\oint \frac{dw}{2\pi i} J^{++}(w) \Gm(z) =
    [J_0^{++},\Gm(z)] \, , \label{eq_GtpDef}
\end{equation}
so that $(\Gp,\Gtm)$ and $(\Gtp,\Gm)$ transform as doublets under $SU(2)$
(or $SU(1,1)$). We will see below, however, that there is a certain freedom
to embed the \N2 superconformal algebra into a small \Ng4 superconformal
algebra, parametrized by $SU(2)$ (or $SU(1,1)$). So, the embedding given above
corresponds to a special choice (see section~\ref{sec_RealProp} for more
details).

This small \Ng4 algebra in general has a nonvanishing central charge. In
topological string theories~\cite{Berkovits:1994vy}, it is removed by 
``twisting'' $T$ by the $U(1)$ current $J$, 
i.\,e.\ $T\to T+\frac{1}{2} \pa J$, so that the
resulting algebra has vanishing central charge. This will be shown explicitly
in the subsequent paragraphs.

\noindent
{\bf Realization in terms of $\bsy{N\,{=}\,1}$ matter multiplets.}
For the construction of an anomaly-free \Ne1 superstring theory, ten massless
matter multiplets are needed. A massless \Ne1 matter multiplet $(X,\psi)$
consists of real bosons $X$ (the ten string coordinates) and $so(1,1)$ Majorana
spinors $\ps$ (their superpartners, each splitting up into a left- and a
right-handed Majorana-Weyl spinor). Due to the reparametrization invariance
of the related supersymmetric sigma model its covariant quantization entails
the introduction of world-sheet (anti)ghosts $b$ and $c$ and their superpartners
$\beta$ and $\gamma$. The superghosts are bosonized in the usual way~%
\cite{Friedan:1985ge}, thereby bringing about the anticommuting fields
$\eta$ and $\xi$. The realization of the previously mentioned \N2
superconformal algebra in terms of these multiplets is given by~%
\cite{Berkovits:1994vy} \\
\parbox{15cm}{
\begin{gather*}
  T = T_{N=1} + \frac{1}{2} \pa (bc+\xi\eta) \, , \\
  \Gm = b\, ,\quad \Gp \, = \, j_{\text{BRST}} + \pa^2 c +
    \pa(c\xi\eta) \, , \\
  J\, = \, cb+\eta\xi \, ,
\end{gather*}}
\hfill
\parbox{8mm}{\begin{eqnarray} \end{eqnarray}} \\
where the energy-momentum tensor $T_{N=1}$ is the usual one with central
charge $c=0$ and $j_{\text{BRST}}$ is the BRST current. These generators make
up an \N2 superconformal algebra with $c=6$, i.\,e.\ with the same central
charge as the critical \N2 superstring. A straightforward method for
calculating scattering amplitudes would be to introduce an \N2 superghost
system with $c^{\text{gh}}=-6$ compensating the positive central charge~%
\cite{Berkovits:1993xq}. However, there is a more elegant method~%
\cite{Berkovits:1994vy}: One can embed the \N2 algebra into a small \Ng4
algebra (as described above) and afterwards twist by the $U(1)$ current
$J$. Then,  \nopagebreak \\
\parbox{15cm}{
\begin{gather*}
  \Gtm = [Q, b \xi]\, = \, -b{\cal X}+\xi T_{N=1} \, ,\quad
    \Gtp \, = \,\eta \, , \\[2mm]
  J^{--} = b\xi \, ,\quad J^{++}\, = \, c\eta\, ,
\end{gather*}}
\hfill
\parbox{8mm}{\begin{eqnarray} \end{eqnarray}} \\
where ${\cal X}$ denotes the picture raising operator and $Q$ is the
BRST-operator of the original \Ne1 string theory. These generators together
make up a small \Ng4 superconformal algebra with $c=6$. The twist~%
\cite{Berkovits:1999im} $T\to T+\frac{1}{2} \pa J$ in this case amounts
to removing the term $\frac{1}{2} \pa (bc+\xi\eta)$ from $T$, thereby
reproducing the original $T_{N=1}$ with $c=0$. It shifts the weight of
each conformal field by $-1/2$ of its $U(1)$ charge --- in particular,
\Gp and \Gtp after twisting become fermionic spin~1 generators which
subsequently serve as BRST-like currents. Their zero modes are exactly
$Q$ and $\eta_0$, respectively.

\noindent
{\bf Realization in terms of $\bsy{N\,{=}\,2}$ matter
multiplets.} For later use, we also discuss the representation in
terms of \N2 multiplets~\cite{Berkovits:1994vy, Junemann:1999hi,
Bellucci:2001qi}: From the world-sheet point of view, critical \N2
strings in flat Kleinian space $\R^{2,2}\cong \C^{1,1}$ are described by
a theory of \N2 supergravity on a (1+1)-dimensional (pseudo)Riemann surface,
coupled to two chiral \N2 massless matter multiplets $(Z,\psi)$. The
components of these multiplets are complex scalars (the four string
coordinates) and $so(1,1)$ Dirac spinors (their four NSR partners). The
latter can be chosen to have unit charge $\pm 1$ under the $U(1)$ current
$J$ of the \N2 superconformal algebra; we denote them by $\ps^\pm$
(restricting ourselves to the chiral half). One has the following
realization of the \N2 super Virasoro algebra in terms of free fields: \\
\parbox{15cm}{
\begin{gather*}
  T = -\frac{1}{4} \eta_{a\ab} \left( 2\pa Z^a\pa\Zb^\ab - \ps^{+a} \pa
    \ps^{-\ab} - \ps^{-\ab} \pa \ps^{+a} \right) \, , \\
  \Gp = \eta_{a\ab} \ps^{+a} \pa \Zb^\ab \, , \qquad \Gm = \eta_{\ab a}
    \ps^{-\ab}\pa Z^a \, , \\
  J = \frac{1}{2} \eta_{\ab a} \ps^{-\ab} \ps^{+a} \, .
\end{gather*}}
\hfill
\parbox{8mm}{\begin{eqnarray} \label{eq_N4SCA_N2_1} \end{eqnarray}} \\
Here $a\in\{0,1\}, \ab\in\{\nb,\ob\}$, and the metric $(\eta_{a\ab})$ with
nonvanishing components $\eta_{1\bar{1}}=-\eta_{0\bar{0}}=1$ defines an
$SU(1,1)$-invariant scalar product. Using the operator product expansions
\begin{equation}
  Z^a(z,\zb) \Zb^\ab (w,\wb) \sim -2 \eta^{a\ab} \ln |z-w|^2 \quad\mbox{and}
    \quad \ps^{+a}(z) \ps^{-\ab}(w) \sim - \frac{2\eta^{a\ab}}{z-w} \, ,
    \label{eq_ZZbOPE}
\end{equation}
it is easy to check that these currents implement the \N2 superconformal
algebra with central charge $c=6$. This central charge is usually
compensated by that of the \N2 (super)ghost Virasoro algebra. Equivalently,
the \N2 super Virasoro algebra can be embedded into a small \Ng4
superconformal algebra~\cite{Berkovits:1994vy} which after twisting
has vanishing central charge. Note that in this approach, contrary to the
\Ne1 case, we do not need to introduce reparametrization ghosts. The
\Ng4 extension is achieved by adding the currents \\
\parbox{15cm}{
\begin{gather*}
  J^{++} = \frac{1}{4}\ve_{ab}\, \ps^{+a}\ps^{+b}\, , \quad J^{--} =
    \frac{1}{4}\ve_{\ab\bb}\, \ps^{-\ab}\ps^{-\bb} \, , \\
  \Gtp = -\ve_{ab}\, \ps^{+a} \pa Z^b \, , \quad \Gtm = -\ve_{\ab\bb}\,
    \ps^{-\ab} \pa\Zb^\bb \, ,
\end{gather*}}
\hfill
\parbox{8mm}{\begin{eqnarray} \label{eq_N4SCA_N2_2} \end{eqnarray}} \\
where we choose the convention that $\ve_{01}=\ve_{\nb\ob}=-\ve^{01}=
-\ve^{\nb\ob}=1$. The twist $T\to T+\frac{1}{2}\pa J$ shifts the weights
by $-1/2$ of their fields' $U(1)$ charge, and again the supercurrents
\Gp and \Gtp become fermionic spin~1 currents (cf.\ the \Ne1 case).

\noindent
{\bf String fields.} In the rest of the paper, we will consider string
fields as functionals of $X^\mu(\s,\t)$, $\ps^\mu(\s,\t),\, b(\s,\t),\,
c(\s,\t),\, \beta(\s,\t)$, and $\gamma(\s,\t)$ for \Ne1 and of $Z^a(\s,\t),
\,\Zb^\ab(\s,\t),\, \ps^{+a}(\s,\t)$, and $\ps^{-\ab}(\s,\t)$ for \N2. We
will often suppress the $\t$-dependence of the world-sheet fields and only
indicate their $\s$-dependence when necessary. As an abbreviation for
this list of world-sheet fields, we use $X$ and $\ps$ and thus denote
a string field $\P$ by $\P[X,\ps]$ (in the \N2 case, one may think of
$X$ as the real and imaginary parts of $Z$, similarly with $\ps$).
Throughout this paper, all string fields are understood to be multiplied
via Witten's star product. Equivalently, one can consider the string fields
as operators (see e.\,g.~\cite{Arefeva:2001ps} and references therein).

\smallskip
\section{Reality properties} \label{sec_RealProp}
\noindent
{\bf Real structures on $\bsy{sl(2,}\Kpmb{\C}\bsy{)}$.} We would like to
introduce a real structure on the Lie algebra $sl(2,\C)$. A real structure
$\s$ on a vector space $V$ is by definition an antilinear involution
$\s\colon V\to V$, i.\,e.\ $\s(\z X+Y) = \bar{\z}\s(X)+\s(Y)$ for
$\z\in\C$, $X,Y \in V$ and $\s^2=\mbox{id}_V$ (here we choose $V=sl(2,\C)$).
There are three choices for a real structure on $sl(2,\C)$. Acting onto
the defining representation of $sl(2,\C)$, we can write them as
\begin{gather}
  \s_\ve \begin{pmatrix} a & b \\ c & -a \end{pmatrix} :=
    \begin{pmatrix} 0 & 1 \\ \ve & 0 \end{pmatrix} \begin{pmatrix}
    \ab & \bb \\ \cb & -\ab \end{pmatrix} \begin{pmatrix} 0 & \ve \\ 1 & 0
    \end{pmatrix} \, \label{eq_realStr_su} \\[1mm]
  \s_0 \begin{pmatrix} a & b \\ c & -a \end{pmatrix} :=
    \begin{pmatrix} \ab & \bb \\ \cb & -\ab \end{pmatrix}
    \label{eq_realStr_sl}
\end{gather}
for $\ve=\pm 1$ and $a,b,c\in \C$. We denote the space of fixed points by
$V_\R$, i.\,e.\ $V_\R:=\{ X\in V: \s(X)=X\}$. For the real structures~%
(\ref{eq_realStr_su}), it is straightforward to check that $V_\R \cong
su(2)$ for $\ve = -1$ and $V_\R \cong su(1,1)$ for $\ve=1$, whereas, for the
real structure $\s_0$, we obtain $V_\R \cong sl(2,\R)$. Real linear
combinations of vectors within $V_\R$ are again contained in $V_\R$, so
they make up real linear subspaces in $sl(2,\C)$.\footnote{In the following,
we will be mainly interested in $\s_\ve$. The case of eq.~(\ref{eq_realStr_sl})
has been used in~\cite{Berkovits:1997pq} and \cite{Lechtenfeld:2000qj}.}
Because of the form of eqs.~(\ref{eq_realStr_su}) and~(\ref{eq_realStr_sl})
it is clear that $\s$ preserves the Lie algebra structure, i.\,e.\ the $V_\R$
are real Lie subalgebras. In the case of $\s_\ve$, complex conjugation can
be ``undone'' by a conjugation with the matrix $\left( \begin{smallmatrix}
0 & 1 \\ \ve & 0 \end{smallmatrix}\right)$ on an element of this linear
subspace. Let us already here note that this conjugation matrix is contained
within the group $SU(2)$ for $\ve=-1$, but is outside $SU(1,1)$ for $\ve=1$.
This means~\cite{Humphreys, Helgason} that conjugation by $\left(
\begin{smallmatrix} 0 & 1 \\ \ve & 0 \end{smallmatrix}\right)$ is an inner
automorphism on $su(2)$ and an outer automorphism on $su(1,1)$.

\noindent
{\bf Bar operation.} As stated above, an \Ng4 superconformal algebra
contains as a subalgebra a current algebra generated by $J$, $J^{++}$, and
$J^{--}$. Its horizontal Lie algebra, when complexified, is $sl(2,\C)$.
On $sl(2,\C)$, let us introduce a real structure $\s_\ve$ from~%
(\ref{eq_realStr_su}), where we choose $\ve=-1$ for \Ne1 and $\ve=1$ for
\N2 strings. The action of complex conjugation on the \Ng4 superconformal
generators, as determined in~\cite{Berkovits:1994vy, Berkovits:1999im},
looks as follows:
\begin{equation}
  J^*=-J\, , \quad (J^{++})^*=J^{--}\, , \quad (\Gp)^*=\Gm\, , \quad
    (\Gtp)^*=\Gtm \, . \label{eq_SCAcc}
\end{equation}
Now, the signflip of $J$ in~(\ref{eq_SCAcc}) necessitates an additional
rotation to reestablish the original twist $T\to T+\frac{1}{2}\pa J$ of
the \Ng4 superconformal algebra. As mentioned in the previous paragraph,
this can be accomplished by an inner automorphism in the case of \Ne1
strings and by an outer automorphism of the current Lie algebra in the
case of \N2 strings. These automorphisms consist in conjugating an element of 
the defining representation with the matrix $\left( \begin{smallmatrix}
0 & 1 \\ \ve & 0 \end{smallmatrix}\right)$. In the following, we want to
scrutinize how the rotation reestablishing the twist acts onto the other
elements of the \Ng4 superconformal algebra.

As explained in~\cite{Berkovits:1999im}, the four spin~$\frac{3}{2}$
superpartners of the energy-momentum tensor transform under the current
group (therefore the current indices $\pm$) as well as under a group%
\footnote{For a discussion of its existence, see~\cite{Berkovits:1999im}.} of
additional automorphisms $SU(2)$ (or $SU(1,1)$). Explicitly,
\begin{equation}
  (G^{\at\a})=\begin{pmatrix}
             \Gp & \Gtp \\ \ve\Gtm & \Gm
          \end{pmatrix} \label{eq_G_matrix}
\end{equation}
transforms under current group transformations by left-multiplication
(note that the columns of this matrix form doublets under the current
algebra, cf.~(\ref{eq_GtmDef}) and~(\ref{eq_GtpDef})). Under additional
automorphisms it transforms by right-multiplication. Obviously, the
rotation reestablishing the original twist acts onto the complex conjugate
matrix as
\begin{equation}
  \begin{pmatrix} 0 & 1 \\ \ve & 0 \end{pmatrix} \begin{pmatrix}
    \Gm & \Gtm \\ \ve \Gtp & \Gp \end{pmatrix} = \begin{pmatrix}
    \ve\Gtp & \Gp \\ \ve\Gm & \ve\Gtm \end{pmatrix} \, . \label{eq_GbarDef}
\end{equation}
Having established the action of this rotation on the \Ng4 superconformal
algebra, we denote the combined operation of complex conjugation and
``twist-restoring'' transformation by a bar, i.\,e.\
\begin{equation}
  \ov{\Gp}=\ve\Gtp\, ,\quad \ov{\Gtp}=\Gp\, ,\quad \ov{\Gtm}=\Gm\, ,\quad
    \mbox{and}\quad \ov{\Gm}=\ve\Gtm\, . \label{eq_BarDef}
\end{equation}
On elements of the complexified current algebra, the bar operation acts
by~(\ref{eq_realStr_su}), in particular it acts trivially onto the real
$su(2)$ or $su(1,1)$ subalgebra.

\noindent
{\bf Additional automorphisms.} The aforementioned additional
automorphisms of the \Ng4 superconformal algebra act on the ``$G$-matrix''
from the right~\cite{Berkovits:1999im}. We take them to be elements of
$SL(2,\C)$ and therefore define
\begin{equation}
  \begin{pmatrix} \Gp(v) & \Gtp(v) \\ \ve\, \Gtm(v) & \Gm(v) \end{pmatrix}
  := \begin{pmatrix} \Gp & \Gtp \\ \ve\,\Gtm & \Gm \end{pmatrix}
  \begin{pmatrix} v_3 & v_1 \\ v_4 & v_2 \end{pmatrix}
\end{equation}
for $v_i\in\C$ with $v_2 v_3-v_1 v_4=1$. The action of the additional
automorphisms should be compatible with the above-defined bar operation~%
(\ref{eq_BarDef}), i.\,e.\
\begin{equation}
  \ov{\Gp(v)}=\ve\Gtp(v)\, ,\quad \ov{\Gtp(v)}=\Gp(v)\, , \quad
  \ov{\Gtm(v)}=\Gm(v)\, , \quad\mbox{and}\quad \ov{\Gm(v)}=\ve\Gtm(v)\, .
  \label{eq_BarComptbl}
\end{equation}
These four equations all lead to the same requirements $v_3=\bar{v}_2=:u_1$
and $v_4=\ve\bar{v}_1=:u_2$, which in effect restricts the group of additional
automorphisms from $SL(2,\C)$ to $SU(2)$ for $\ve=-1$ and to $SU(1,1)$ for
$\ve=1$:
\begin{equation}
  \begin{pmatrix} \Gp(u) & \Gtp(u) \\ \ve\, \Gtm(u) & \Gm(u) \end{pmatrix}
  := \begin{pmatrix} \Gp & \Gtp \\ \ve\,\Gtm & \Gm \end{pmatrix}
  \begin{pmatrix} u_1 & \ve\ub_2 \\ u_2 & \ub_1 \end{pmatrix}
  \label{eq_GuDef}
\end{equation}
with $|u_1|^2-\ve |u_2|^2=1$. Note that
\begin{equation}
  \begin{pmatrix} 0 & \ve \\ 1 & 0\end{pmatrix} \begin{pmatrix} u_1 &
    \ve \ub_2\\ u_2 & \ub_1\end{pmatrix} \begin{pmatrix} 0 & 1 \\ \ve & 0
    \end{pmatrix} = \begin{pmatrix} \ub_1 & \ve u_2 \\ \ub_2 & u_1
    \end{pmatrix}\, . \label{eq_uReal}
\end{equation}
Obviously, eq.~(\ref{eq_BarComptbl}) entails the introduction of the same
real structure on the group of additional automorphisms as the one chosen
for the current group. Since we are only interested in the ratio of the
prefactors of \Gp and $\Gtp$, and of \Gm and $\Gtm$, respectively, we define
for later use the combinations\\
\parbox{15cm}{
\begin{eqnarray*}
  \Gtp(\l) & := & \Gtp + \l \Gp \, = \, \frac{1}{\ub_1}\, \Gtp(u) \, , \\
  \Gtm(\l) & := & \Gtm + \lb \Gm \, = \, \frac{1}{u_1}\, \Gtm(u) \, , \\
  \Gp(\l) & := & \Gp + \ve\lb\Gtp \, = \, \frac{1}{u_1}\, \Gp(u) \, , \\
  \Gm(\l) & := & \Gm + \ve\l\Gtm \, = \, \frac{1}{\ub_1}\, \Gm(u) \, .
\end{eqnarray*}}
\hfill
\parbox{8mm}{\begin{eqnarray} \label{eq_GlDef} \end{eqnarray}} \\
Here, $(\ub_1:\ve\ub_2)$ can be regarded as homogeneous coordinates on the
sphere $S^2\cong \C P^1$, and $\l\equiv \ve \ub_2/\ub_1$ is a local coordinate
for $\ub_1\neq 0$.

Let us reconsider the action of the bar operation on the 
matrix~(\ref{eq_GuDef}).
Remember that the bar operation consists of a complex conjugation and a
subsequent twist-restoring rotation as in~(\ref{eq_GbarDef}). Acting
on~(\ref{eq_GuDef}), we have \\
\parbox{15cm}{
\begin{eqnarray*}
  \begin{pmatrix} 0 & 1 \\ \ve & 0 \end{pmatrix} \begin{pmatrix}
    \Gm & \Gtm \\ \ve \Gtp & \Gp \end{pmatrix} \begin{pmatrix}
    \ub_1 & \ve u_2 \\ \ub_2 & u_1 \end{pmatrix} & = & \begin{pmatrix}
    \ve\Gtp & \Gp \\ \ve\Gm & \ve\Gtm \end{pmatrix} \begin{pmatrix}
    0 & \ve \\ 1 & 0\end{pmatrix} \begin{pmatrix}u_1 & \ve \ub_2 \\
    u_2 & \ub_1 \end{pmatrix} \begin{pmatrix} 0 & 1 \\ \ve & 0
    \end{pmatrix} \\
    & = & \begin{pmatrix} \Gp & \Gtp \\ \ve\Gtm & \Gm \end{pmatrix}
    \begin{pmatrix} \ub_2 & u_1 \\ \ve\ub_1 & u_2 \end{pmatrix}\, .
\end{eqnarray*}}
\hfill
\parbox{8mm}{\begin{eqnarray} \end{eqnarray}} \\
For the first equality, we have used~(\ref{eq_GbarDef}) and~(\ref{eq_uReal}).
An additional right-multiplication by $\left( \begin{smallmatrix} 0 & \ve \\
1 & 0 \end{smallmatrix}\right)$ transforms the ``$u$-matrix'' on the right
hand side back to its original form (cf.~(\ref{eq_uReal})), thereby taking
$\ve\lb^{-1}$ to $\l$. This mediates a map between operators defined for
$|\l|<\infty$ and operators defined for $|\l|>0$.

The action of the bar operation on $\Gtp(\ve\lb^{-1})$ etc.\ can
be determined from the fact that $\ve\lb^{-1}=u_1/u_2$ and
therefore
\begin{equation}
  \begin{pmatrix} \Gp(\ve\lb^{-1}) & \Gtp(\ve\lb^{-1}) \\
    \ve\Gtm(\ve\lb^{-1}) & \Gm(\ve\lb^{-1}) \end{pmatrix} = \begin{pmatrix}
    \Gp & \Gtp \\ \ve\Gtm & \Gm \end{pmatrix} \begin{pmatrix} \ub_2 & u_1 \\
    \ve\ub_1 & u_2 \end{pmatrix} \begin{pmatrix} \ub_2^{-1} & 0 \\ 0 & u_2^{-1}
    \end{pmatrix} \, .
\end{equation}
Multiplying the complex conjugate matrices from the left by a twist-restoring
rotation, we obtain
\begin{equation}
  \begin{pmatrix} 0 & 1 \\ \ve & 0 \end{pmatrix} \begin{pmatrix}
    \Gm & \Gtm \\ \ve \Gtp & \Gp \end{pmatrix} \begin{pmatrix}
    u_2 & \ub_1 \\ \ve u_1 & \ub_2 \end{pmatrix} \begin{pmatrix}
    u_2^{-1} & 0 \\ 0 & \ub_2^{-1} \end{pmatrix} = \begin{pmatrix}
    \lb^{-1}\Gp(\l) & \l^{-1}\Gtp(\l) \\ \ve\lb^{-1}\Gtm(\l)
    & \l^{-1}\Gm(\l) \end{pmatrix}\, .
\end{equation}
So, ``barring'' accompanied by the transformation $\l\mapsto\ve\lb^{-1}$
maps $\Gtp(\l)$ (defined for $|\l|<\infty$) to $\frac{1}{\l}\,\Gtp(\l)$
(defined for $|\l|>0$),
\begin{equation}
  \ov{\Gtp(\ve\lb^{-1})} = \frac{1}{\l}\,\Gtp(\l)\, .
\end{equation}
This result will be needed in section~\ref{sec_Integr}.

For the selection of an \N2 superconformal subalgebra within the \Ng4
algebra, there is obviously the freedom to choose a linear combination
of the $J$'s as the $U(1)$ current. All choices are equivalent
through current $SU(2)$- (or $SU(1,1)$-) rotations (acting on the
matrices in~(\ref{eq_G_matrix}) from the left). In addition, there is
the freedom to choose which linear combination of the positively charged
$G$'s will be called~$G^+$~\cite{D'Adda:1987rx}. This freedom is
parametrized by another $SU(2)$ (or $SU(1,1)$), cf.~(\ref{eq_GuDef}).
Since in our case only the ratio of the prefactors of \Gp and \Gtp is
important, we arrive at generators~(\ref{eq_GlDef}) parametrized by
$\l\in\C P^1$. So, we obtain a one-parameter family of \N2 superconformal
algebras embedded into a small \Ng4 algebra~\cite{Berkovits:1994vy,
Berkovits:1999im}.

\smallskip
\section{Integrability of Berkovits' string field theory} \label{sec_Integr}
\noindent
In this section we show that the equation of motion~(\ref{eq_BSSFTeom}) of
Berkovits' string field theory can be obtained as the compatibility
condition of some linear equations. In other words, solutions of these
linear equations exist iff eq.~(\ref{eq_BSSFTeom}) is satisfied. For \N2
strings the integrability of Berkovits' string field theory was shown in~%
\cite{Lechtenfeld:2000qj}, and here we extend this analysis to the \Ne1 case.
The payoff for considering such (integrable) models is the availability
of powerful techniques for the construction of solutions to the equation(s)
of motion.

\noindent
{\bf Sphere $\Kpmb{\C}\bsy{P^1}$.} Let us consider the Riemann sphere
$S^2\cong \C P^1 \cong \C\cup\{\infty\}$ and cover it by two coordinate
patches
\begin{equation}
  \C P^1 = U_+ \cup U_-\, , \quad U_+ := \{ \l\in\C : |\l|<1+\eps\} \, ,
    \quad U_- := \{ \l\in\C\cup\{\infty\}: |\l|>(1+\eps)^{-1}\}
\end{equation}
for some $\eps>0$ with the overlap
\begin{equation}
  U_+\cap U_- \supset S^1 = \{ \l\in\C : |\l| = 1\} \, .
\end{equation}
We will consider $\l\in U_+$ and $\lt\in U_-$ as local complex coordinates
on $\C P^1$ with $\lt= \frac{1}{\l}$ in $U_+\cap U_-$. Recall that $\l\in
\C P^1$ was introduced in the previous section as a parameter for a family
of \N2 subalgebras in the twisted small \Ng4 superconformal algebra with
fermionic currents~(\ref{eq_GlDef}).

\noindent
{\bf Linear system.} Taking the string field $\P$ from~(\ref{eq_BSSFTact})
and operators $\Gtp(\l)$ from~(\ref{eq_GlDef}), we introduce the following
equation:
\begin{equation}
  (\Gtp + \l \Gp + \l A) \Psi = 0 \, , \label{eq_PsiLin}
\end{equation}
where $A:=e^{-\P}\Gp e^\P$ and $\Psi$ is a matrix-valued string field
depending not only on $X$ and $\ps$ but also (meromorphically) on the
auxiliary parameter $\l\in\C P^1$. As in eq.~(\ref{eq_opact}) the action
of \Gp and \Gtp on $\Psi$ implies a contour integral of the (fermionic)
current around $\Psi$. Note that \Gp and \Gtp are Grassmann-odd and
therefore $(\Gp)^2 = (\Gtp)^2 = \Gp\Gtp + \Gtp\Gp = 0$.

If $e^\P$ is given, then~(\ref{eq_PsiLin}) is an equation for the field
$\Psi$. Solutions $\Psi$ of this linear equation exist if the term in
brackets squares to zero, i.\,e.\
\begin{equation}
  \left( \Gtp + \l\Gp + \l A \right)^2=0 \quad \Leftrightarrow \quad
    \l^2(\Gp + A)^2 + \l \Gtp A=0
\end{equation}
for any $\l$. Here we have used the Grassmann nature of \Gp and \Gtp.
So, we obtain two equations:
\begin{eqnarray}
  \Gp A+A^2 & = & 0 \, , \\
  \Gtp A & = & 0\, .
\end{eqnarray}
The first of these two equations is trivial since \Gp acts as a derivation
on the algebra of string fields. The second equation coincides with the
equation of motion~(\ref{eq_BSSFTeom}).

\noindent
{\bf Chiral string fields.} As a special case of eq.~(\ref{eq_PsiLin}) one
can consider the equations
\begin{eqnarray}
  \Big(\Gtp + \l \Gp + \l A \Big) \Psi_+  & =  & 0 \, , \label{eq_linPsip} \\
  \Big( \frac{1}{\l}\, \Gtp + \Gp + A \Big) \Psi_- & = & 0\, ,
    \label{eq_linPsim}
\end{eqnarray}
where $\Psi_+$ and $\Psi_-$ are invertible matrix-valued string fields
depending holomorphically on $\l$ and $\frac{1}{\l}$,
respectively.

Considering $\l\to 0$ in~(\ref{eq_linPsip}), we see that $\Gtp \Psi_+(\l=0)
= 0$, and one may choose $\Psi_+(\l=0) = {\cal I}$. Analogously, taking
$\l\to\infty$ in~(\ref{eq_linPsim}), we obtain
\begin{equation}
  A = e^{-\P}\Gp e^\P = \Psi_-(\l=\infty) \Gp \Psi_-^{-1}(\l=\infty) \, ,
\end{equation}
and one may choose $\Psi_-(\l=\infty) = e^{-\P}$ as a solution thereof.
From this we derive the asymptotic behavior of our fields:
\begin{gather}
  \Psi_+ = {\cal I} + O(\l) \quad \mbox{for }\l\to 0\, , \label{eq_PsipAsympt} \\
  \Psi_- = e^{-\P} + O(\l^{-1}) \quad \mbox{for }\l\to\infty\, .
    \label{eq_PsimAsympt}
\end{gather}
We see that $\Psi_-$ may be considered as an $\l$-augmented solution of eq.~%
(\ref{eq_BSSFTeom}), and all information about $e^\P$ is contained in
$\Psi_\pm$.

Suppose that we find solutions $\Psi_+$ and $\Psi_-$ of eqs.~(\ref{eq_linPsip})
and~(\ref{eq_linPsim}) for a given $e^\P$. Then one can introduce the matrix-%
valued string field
\begin{equation}
  \Upsilon_{+-} := \Psi_+^{-1} \Psi_-
\end{equation}
defined for $\l\in U_+\cap U_-$. From eqs.~(\ref{eq_linPsip}) and~%
(\ref{eq_linPsim}) it follows that
\begin{equation}
  \Gtp(\l) \Upsilon_{+-} \equiv (\Gtp + \l \Gp) \Upsilon_{+-} = 0 \, .
    \label{eq_TransFunc}
\end{equation}
String fields annihilated by the operator $\Gtp(\l)$ will be called
{\em chiral}.

\noindent
{\bf Reality properties.} W.\,r.\,t.\ the bar operation from section~%
\ref{sec_RealProp}, the string field $\P$ is real\footnote{The world-sheet
parity transformation $\s\mapsto \pi-\s$ is accompanied by a transposition
of Chan-Paton matrices. Note that hermitian generators are used for the
$u(n)$ Chan-Paton algebra. Therefore, $e^\P$ does not necessarily take
values in $U(n)$.}, i.\,e.~%
\cite{Berkovits:1995ab}
\begin{equation}
  \ov{\P[X(\pi-\s),\ps(\pi-\s)]} = \P[X(\s),\ps(\s)] \, .
\end{equation}
To see the behavior of the extended string field under the bar operation
we scrutinize eqs.~(\ref{eq_linPsip}) and~(\ref{eq_linPsim}). We already
saw that $\ov{\Gtp(\ve\lb^{-1})} = \frac{1}{\l} \Gtp + \Gp$, and by definition,
$\ov{A} = (\ov{\Gp} e^{\ov{\P}}) e^{-\ov{\P}} = \ve (\Gtp e^\P) e^{-\P}$
under $\s\mapsto\pi-\s$. Then mapping $\l\mapsto\ve\lb^{-1}$ and $\s\mapsto
\pi-\s$ in~(\ref{eq_linPsip}) and conjugating, we obtain
\begin{equation}
  \Big( \frac{1}{\l}\,\Gtp + \Gp \Big) \Psb_+^{-1} - \frac{1}{\l} (\Gtp e^\P)
    e^{-\P} \Psb_+^{-1} = 0\, .
\end{equation}
This coincides with eq.~(\ref{eq_linPsim}) if we set
\begin{equation}
  \left( \Psb_+\right)^{-1}\Big[ X(\pi-\s),\ps(\pi-\s),\frac{\ve}{\lb} \Big]
  = e^\P \Psi_-\Big[X(\s),\ps(\s),\l \Big] \, . \label{eq_PsipPsimrel}
\end{equation}
Then from~(\ref{eq_PsipPsimrel}) it follows that
\begin{equation}
  \Upsilon_{+-} = \Psb_- e^\P \Psi_-
\end{equation}
is real, i.\,e.\ $\ov{\Upsilon}_{+-}\big[ X(\pi-\s),\ps(\pi-\s),\ve\lb^{-1}
\big] = \Upsilon_{+-}\big[X(\s),\ps(\s),\l \big]$.

\noindent
{\bf Gauge freedom.} Recall that gauge transformations of the fields $e^\P$ and
$A$ have the form \\
\parbox{15cm}{
\begin{gather*}
  e^\P \mapsto e^{\P'} = B e^\P C \quad\mbox{with}\quad \Gp B = 0\, ,
    \quad \Gtp C = 0\, , \\
  A \mapsto A' = C^{-1} A C + C^{-1} \Gp C\, .
\end{gather*}}
\hfill
\parbox{8mm}{\begin{eqnarray} \label{eq_gaugeinv} \end{eqnarray}} \\
Under $C$-transformations the fields $\Psi_\pm$ transform as
\begin{equation}
  \Psi_\pm \mapsto \Psi'_\pm = C^{-1} \Psi_\pm\, . \label{eq_PsiInvar}
\end{equation}
It is easy to see that the chiral string field $\Upsilon_{+-}$ is invariant
under the transformations~(\ref{eq_PsiInvar}). On the other hand, the field
$A$ will remain unchanged after the transformations
\begin{equation}
  \Psi_+ \mapsto \Psi_+ h_+\, , \quad \Psi_- \mapsto \Psi_- h_- \, ,
    \label{eq_hpmgauge}
\end{equation}
where $h_+$ and $h_-$ are chiral string fields depending holomorphically
on $\l$ and $\frac{1}{\l}$, respectively. In the special case when $h_+$
and $h_-$ are independent of $\l$ the transformations~(\ref{eq_hpmgauge})
with $h_+ = {\cal I}$, $h_- =: B^{-1}$ induce the $B$-transformations $e^\P
\mapsto B e^\P$ in~(\ref{eq_gaugeinv}). In general, eqs.~(\ref{eq_hpmgauge})
induce the transformations
\begin{equation}
  \Upsilon_{+-} \mapsto h_+^{-1} \Upsilon_{+-} h_- \label{eq_UpsTrans}
\end{equation}
on the space of solutions to eq.~(\ref{eq_TransFunc}), and any two solutions
differing by a transformation~(\ref{eq_UpsTrans}) are considered to be
equivalent.

\noindent
{\bf Splitting.} Up to now, we discussed how to find $\Upsilon_{+-}$ for
a given $A=e^{-\P}\Gp e^\P$. Consider now the converse situation. Suppose
we found a string field $\Upsilon_{+-}$ which depends analytically on $\l\in
S^1$ and satisfies the linear equation~(\ref{eq_TransFunc}). Being real-%
analytic $\Upsilon_{+-}$ can be extended to a string field depending
holomorphically on $\l\in U_+\cap U_-$. Then we can formulate an operator
version of the Riemann-Hilbert problem: Split $\Upsilon_{+-}=\Psi_+^{-1}
\Psi_-$ into matrix-valued string fields $\Psi_\pm$ depending on
$\l\in U_+\cap U_-$ such that $\Psi_+$ can be extended to a regular
(i.\,e.\ holomorphic in $\l$ and invertible) matrix-valued function on
$U_+$ and that $\Psi_-$ can be extended to a regular matrix-valued
function on $U_-$. From eq.~(\ref{eq_TransFunc}) it then follows that
\begin{equation}
  \Psi_+ (\Gtp + \l\Gp) \Psi_+^{-1} = \Psi_-(\Gtp + \l\Gp) \Psi_-^{-1} =
    \tilde{A} + \l A \, ,
\end{equation}
where $\tilde{A}$ and $A$ are some $\l$-independent string fields. The last
equality follows from expanding $\Psi_+$ and $\Psi_-$ into power series in
$\l$ and $\l^{-1}$, respectively. If we now choose a function $\Psi_+(\l)$
such that\footnote{This can always be achieved by redefining $\Psi_+(\l)
\mapsto \Psi_+^{-1}(\l=0)\Psi_+(\l)$, $\Psi_-(\l^{-1})\mapsto \Psi_+^{-1}
(\l=0)\Psi_-(\l^{-1})$.} $\Psi_+(\l=0) = {\cal I}$ then $\tilde{A}=0$ and
\begin{equation}
  A = \Psi_-(\l=\infty)\Gp\Psi_-^{-1}(\l=\infty) \quad \Rightarrow \quad
    \Psi_-(\l=\infty) =: e^{-\P} \, .
\end{equation}
In the general case, we have
\begin{equation}
  e^{-\P} = \Psi_+^{-1}(\l=0) \Psi_-(\l=\infty)\, .
\end{equation}
So, starting from $\Upsilon_{+-}$ we have constructed a solution $e^\P$ of
eq.~(\ref{eq_BSSFTeom}).

Suppose that we know a splitting for a given $\Upsilon_{+-}$ and have
determined
a correspondence $\Upsilon_{+-} \leftrightarrow e^\P$. Then for any matrix-%
valued chiral string field $\widetilde{\Upsilon}_{+-}$ from a small enough
neighborhood of $\Upsilon_{+-}$ (i.\,e.\ $\widetilde{\Upsilon}_{+-}$ is close
to $\Upsilon_{+-}$ in some norm) there exists a splitting $\widetilde{
\Upsilon}_{+-}=\widetilde{\Psi}_+^{-1}\widetilde{\Psi}_-$ due to general
deformation theory arguments. Namely, there are no obstructions to a
deformation of a trivial holomorphic vector bundle $\mathcal{E}$ over $\C P^1$
since its infinitesimal deformations are parametrized by the group $H^1
(\C P^1,\mathcal{E})$. This cohomology group is trivial because of $H^1
(\C P^1,{\cal O})=0$ where ${\cal O}$ is the sheaf of holomorphic functions
on $\C P^1$. But from the correspondence $\widetilde{\Upsilon}_{+-}
\leftrightarrow e^{\widetilde{\P}}$ it follows that any solution
$e^{\widetilde{\P}}$ from an open neighborhood (in the solution space) of
a given solution $e^\P$ can be obtained from a ``free'' chiral string field
$\widetilde{\Upsilon}_{+-}$. In this sense, Berkovits' string field theory
is an integrable theory; in other words, it is completely solvable.

To sum up, we have described a one-to-one correspondence between the gauge
equivalence classes of solutions to the nonlinear equation of
motion~(\ref{eq_BSSFTeom})
and equivalence classes of solutions (chiral string fields) to the auxiliary
linear equation~(\ref{eq_TransFunc}). The next step is to show how this
correspondence helps to solve~(\ref{eq_BSSFTeom}).

\smallskip
\section{Exact solutions by the splitting approach} \label{sec_SplitAppr}
\noindent
{\bf Atiyah-Ward ansatz.} As described in the previous section, solutions
of the string field equations~(\ref{eq_BSSFTeom}) can be obtained by
splitting a given matrix-valued chiral string functional $\Upsilon_{+-}$.
In general, splitting is a difficult problem but for a large class of
special cases it can be achieved. The well known cases (for $n=2$) are
described e.\,g.\ by the infinite hierarchy of Atiyah-Ward ans\"atze~%
\cite{Atiyah:pw} generating instantons in four-dimensional $SU(2)$ Yang-%
Mills theory. These ans\"atze are easily generalized~\cite{Lechtenfeld:2001ie}
to the case of noncommutative instantons, the first examples of which were
given in~\cite{Nekrasov:1998ss}. Here, we consider the first Atiyah-Ward
ansatz from the above-mentioned hierarchy~\cite{Atiyah:pw} and discuss
its generalization to the string field case.

We start from the $2\times 2$ matrix
\begin{equation}
  \Upsilon_{+-} = \begin{pmatrix} \r & \l^{-1} {\cal I} \\
    \ve\l {\cal I} & 0 \end{pmatrix} \, , \label{eq_Upspm_rho}
\end{equation}
where $\r$ is a real and chiral string field, i.\,e.\
\begin{equation}
  \ov{\r}\left[ X,\ps,\frac{\ve}{\lb}\right] = \r[X,\ps,\l]
  \label{eq_rhoreal}
\end{equation}
and
\begin{equation}
  (\Gtp + \l\Gp)\r = 0 \, . \label{eq_rhochiral}
\end{equation}
We assume that $\r$ depends on $\l\in S^1$ analytically and therefore can
be extended holomorphically in $\l$ to an open neighborhood $U_+\cap U_-$
of $S^1$ in $\C P^1$. From~(\ref{eq_rhoreal}) and~(\ref{eq_rhochiral}) it
follows that the matrix $\Upsilon_{+-}$ in~(\ref{eq_Upspm_rho}) is chiral
and real.

\noindent
{\bf Splitting.} We now expand $\r$ into a Laurent series in $\l$,
\begin{equation}
  \r = \sum_{k=-\infty}^\infty \l^k \r_k = \r_- + \r_0 + \r_+ \, , \quad
  \r_- = \sum_{k<0} \l^k \r_k\, , \quad \r_+ = \sum_{k>0} \l^k \r_k \, ,
  \label{eq_rhoexp}
\end{equation}
and obtain from~(\ref{eq_rhochiral}) for
\begin{equation}
  \r_k = \oint \frac{d\l}{2\pi i} \l^{-k-1}\r
\end{equation}
the recursion relations
\begin{equation}
  \Gtp\r_{k+1} = - \Gp\r_k \, . \label{eq_rkRecRel}
\end{equation}
Using~(\ref{eq_rhoexp}), one easily checks that
\begin{equation}
  \Upsilon_{+-} = \Psh_+^{-1} \Psh_- \label{eq_PshDef}
\end{equation}
where
\begin{equation}
  \Psh_+^{-1} = \begin{pmatrix} \r_0+\r_+ & -\l^{-1} \r_+ \\
    \ve\l {\cal I} & -\ve {\cal I} \end{pmatrix}
    \r_0^{-1/2}\, , \qquad \Psh_- = \r_0^{-1/2}\begin{pmatrix} \r_0+\r_- &
    \l^{-1} {\cal I} \\ \l\r_- & {\cal I} \end{pmatrix} \, .
\end{equation}
However, the asymptotic value of $\Psh_+$,
\begin{equation}
  \Psh_+(\l=0) = \r_0^{-1/2} \left. \begin{pmatrix} {\cal I} &
    -\ve\l^{-1}\r_+ \\
    \l {\cal I} & -\ve(\r_0+\r_+) \end{pmatrix}  \right|_{\l=0} =
      \r_0^{-1/2}
  \begin{pmatrix} {\cal I} & -\ve\r_1 \\ 0 & -\ve\r_0 \end{pmatrix}
    \neq \bsy{1}_2 {\cal I} \, ,
\end{equation}
shows that this splitting corresponds to a more general gauge than the
one used in eq.~(\ref{eq_BSSFTeom})
(see~\cite{Berkovits:1997pq,Lechtenfeld:2000qj} for a
discussion of this gauge in the case of \N2 strings).

To obtain the asymptotic behavior~(\ref{eq_PsipAsympt}) one may exploit
the ``gauge freedom'' contained in~(\ref{eq_PshDef}) and introduce the
fields
\begin{equation}
  \Psi_+ := \Psh_+^{-1}(\l=0) \Psh_+ \quad\mbox{and}\quad \Psi_- :=
    \Psh_+^{-1}(\l=0) \Psh_-
\end{equation}
which by definition have the right asymptotic behavior. These functionals
yield the same chiral string field since
\begin{equation}
  \Upsilon_{+-} = \Psh_+^{-1} \Psh_- = \Psh_+^{-1} \Psh_+(\l=0)
    \Psh_+^{-1}(\l=0) \Psh_- = \Psi_+^{-1} \Psi_- \, .
\end{equation}

\noindent
{\bf Explicit solutions.} Now, from
\begin{equation}
  \Psi_- =
  \begin{pmatrix} \r_0+\r_- -\l\r_1\r_0^{-1}\r_- & \l^{-1} {\cal I}
    -\r_1\r_0^{-1} \\
    -\ve\l\r_0^{-1}\r_- & -\ve\r_0^{-1} \end{pmatrix}
\end{equation}
we can determine a solution of~(\ref{eq_BSSFTeom}) with the help
of~(\ref{eq_PsimAsympt}),
\begin{equation}
  e^{-\P} = \Psi_-(\l=\infty) = \begin{pmatrix}
    \r_0-\r_1\r_0^{-1}\r_{-1} & -\r_1\r_0^{-1} \\
    -\ve\r_0^{-1}\r_{-1} & -\ve\r_0^{-1}
  \end{pmatrix} \, . \label{eq_splitsol}
\end{equation}
A direct calculation shows that this satisfies eq.~(\ref{eq_BSSFTeom})
iff $\r_0$, $\r_1$ and $\r_{-1}$ satisfy the linear recursion relations~%
(\ref{eq_rkRecRel}) for $k=-1,0$. Moreover, substituting~(\ref{eq_splitsol})
into~(\ref{eq_BSSFTeom}) yields
\begin{equation}
  \Gtp \Gp \r_0 = 0 \label{eq_KG}
\end{equation}
which is the analogue of the Laplace equation in the case of instantons in
four-dimensional Euclidean space~\cite{Lechtenfeld:2001ie, Nekrasov:1998ss}.
Note that~(\ref{eq_KG}) is just one of an infinite set of equations,
\begin{equation}
  \Gtp\Gp \r_k = 0 \qquad \forall k\in\Z \, ,
\end{equation}
which can easily be obtained from the recursion relations~(\ref{eq_rkRecRel}).
So, the ansatz~(\ref{eq_Upspm_rho}) for $\Upsilon_{+-}$ and its splitting
reduce the nonlinear string field theory equation~(\ref{eq_BSSFTeom}) to
the linear equations~(\ref{eq_rkRecRel}) which are equivalent to the
chirality equation~(\ref{eq_rhochiral}).\footnote{Considering~%
(\ref{eq_splitsol}) as an ansatz, we can obviously relax the chirality
condition~(\ref{eq_rhochiral}) and substitute it with the demand that~%
(\ref{eq_rkRecRel}) is satisfied for $k=-1,0$. For trivial $\Gp$- and
$\Gtp$-cohomology, however, this is again equivalent to eq.~%
(\ref{eq_rkRecRel}) for all $k\in\Z$.}

Finally, notice that in the case of \Ne1 strings, one may take $|\r_0
\rangle = \xi_0 |V\rangle$ where $|V\rangle$ is a state in the ``small''
Hilbert space~\cite{Friedan:1985ge}. Then eq.~(\ref{eq_KG}) reduces to
$Q\eta_0 (\xi_0 |V\rangle) = Q(-\xi_0\eta_0 |V\rangle + |V\rangle)=0$,
i.\,e.\
\begin{equation}
  Q|V\rangle=0 \qquad\mbox{and}\qquad \eta_0 |V\rangle=0\, .
\end{equation}
This fits in nicely with the discussion in~\cite{Berkovits:2001nr}.

\smallskip
\section{Exact solutions via dressing of a seed solution} \label{sec_DressAppr}
\noindent
{\bf Extended solutions.} In the previous section we discussed solutions
$\Psi_+$ and $\Psi_-$ of the linear system which are holomorphic in $\l$
and $\frac{1}{\l}$, respectively. Now we are interested in those solutions
$\Psi$ of eq.~(\ref{eq_PsiLin}) which are holomorphic in open neighborhoods
of both $\l=0$ and $\l=\infty$ and therefore have poles\footnote{By
Liouville's theorem there are no globally defined holomorphic functions on
$\C P^1$ besides constants.} at finite points $\l=\mu_k$, $k=1,\ldots,m$.
Again we see from~(\ref{eq_PsiLin}) that $\Psi(\l=\infty)$ coincides with
$e^{-\P}$ up to a gauge transformation and we fix the gauge by putting
\begin{equation}
  \Psi^{-1}(\l=\infty) = e^\P\, . \label{eq_PsiInf}
\end{equation}
The string field $\Psi[X,\ps,\l]$ will be called the {\it extended solution}
corresponding to $e^\P$. Recall that $e^\P$ is a 
solution of~(\ref{eq_BSSFTeom}) where $\P$ carries $u(n)$ Chan-Paton labels.

The reality properties of extended solutions are derived in very much the
same way as the reality properties of $\Psi_+$ and $\Psi_-$. Namely, one
can easily show that if $\Psi[X,\ps,\l]$ satisfies eq.~(\ref{eq_PsiLin})
then $e^{-\P}\Psb^{-1}\left[X,\ps,\ve\lb^{-1}\right]$ satisfies the
same equation and therefore
\begin{equation}
  \ov{\Psi\Big[ X(\pi-\s,\t),\psi(\pi-\s,\t),\frac{\ve}{\bar{\l}}
    \Big]} =\Big[ \Psi\left[ X(\s,\t),\psi(\s,\t),\l
    \right] \Big]^{-1} e^{-\P} \, , \label{eq_hermPsiPsidag2}
\end{equation}
or equivalently,
\begin{equation}
  \Psi\big[ X(\s,\t),\psi(\s,\t),\l \big] \ov{\Psi\big[
    X(\pi-\s,\t),\psi(\pi-\s,\t),\frac{\ve}{\bar{\l}} \big]}
    = e^{-\P} \, . \label{eq_hermPsiPsidag}
\end{equation}

Using~(\ref{eq_hermPsiPsidag}), one can rewrite eq.~(\ref{eq_PsiLin}) in
the form
\begin{equation}
  \left[ \left(\frac{1}{\l}\Gtp + \Gp \right) \Psi\left[ X,\psi,\l\right]
  \right] \ov{\Psi\left[ X,\psi,\frac{\ve}{\bar{\l}}\right] }=-A\, e^{-\P} \, .
    \label{eq_DrApprLin2}
\end{equation}
Notice that $\Psi$ satisfies the same equation as $\Psi_-$, and therefore
$\Xi:=\Psi^{-1}\Psi_-$ is annihilated by the operator $\Gtp(\l)$. Thus,
\begin{equation}
  \Psi_- = \Psi\Xi \quad\Rightarrow\quad \Psi_+^{-1} = \Psb_- e^\P =
    \ov{\Xi}\, \Psb e^\P \label{eq_OmRel}
\end{equation}
for some matrix-valued chiral string field $\Xi$. Moreover, from~%
(\ref{eq_hermPsiPsidag2}) and~(\ref{eq_OmRel}) we see that
\begin{equation}
  \Upsilon_{+-} = \Psi_+^{-1} \Psi_- = \ov{\Xi}\, \Psb e^\P \Psi \Xi
  = \ov{\Xi}\, \Xi \, .
\end{equation}
This establishes a connection with the discussion in
section~\ref{sec_SplitAppr}.

\noindent
{\bf Dressing.} The dressing method is a recursive procedure generating
a new extended solution from an old one. A solution $e^\P$ of the equation
of motion~(\ref{eq_BSSFTeom}) is obtained from the extended solution via~%
(\ref{eq_PsiInf}). Namely, let us suppose that we have constructed an
extended seed solution $\Psi_0$ by solving the linear equation~%
(\ref{eq_PsiLin}) for a given (seed) solution $e^{\P_0}$ of
eq.~(\ref{eq_BSSFTeom}). Then one can look for a new extended solution in
the form
\begin{equation}
  \Psi_1 = \chi_1 \Psi_0 \qquad\mbox{with}\qquad \chi_1 = {\cal I} +
    \frac{\l\a_1}{\l-\mu_1}\, P_1\, , \label{eq_DressGen1}
\end{equation}
where $\a_1$ and $\mu_1$ are complex constants and the matrix-valued string
field $P_1[X,\ps]$ is independent of $\l$. The transformation $\Psi_0
\mapsto \Psi_1$ is called {\em dressing}. Below, we will show explicitly how
one can determine $\Psi_1$ by exploiting the pole structure (in $\l$)
of eq.~(\ref{eq_DrApprLin2}) together with~(\ref{eq_DressGen1}). An $m$-%
fold repetition of this procedure yields as the new extended solution
\begin{equation}
  \Psi_m = \prod_{j=1}^m \left( {\cal I} + \frac{\l\a_j}{\l-\mu_j}\, P_j
    \right) \Psi_0\, . \label{eq_DressGen2}
\end{equation}
We will choose below the vacuum seed solution $\P_0=0, \Psi_0={\cal I}$.

\noindent
{\bf First-order pole ansatz for $\bsy{\Psi}$.} Choose the complex
constants $\mu_j$ in~(\ref{eq_DressGen2}) such that they are mutually
different. Then using a decomposition into partial fractions, one can
rewrite the multiplicative ansatz~(\ref{eq_DressGen2}) in the additive
form
\begin{equation}
  \Psi_m = \Big( {\cal I} +\l \sum_{q=1}^m \frac{R_q}{\l-\mu_q} \Big)
    \Psi_0\, , \label{eq_psians}
\end{equation}
where the matrix-valued string fields $R_q[X,\ps]$ are some combinations
of (products of) $P_j$. As already mentioned we now choose the vacuum
$\P_0=0$,$\Psi_0={\cal I}$ and consider $R_q$ of the form~\cite{Zakharov:1979,
Forgacs:1983gr, Ward:vc, Lechtenfeld:2001aw}
\begin{equation}
  R_q = -\sum_{p=1}^m \mu_q T_p \G^{pq} \Tb_q \, , \label{eq_RkDef}
\end{equation}
where $T_p[X,\psi]$ are taken to be the $n\times r$ matrices for some
$r\geq 1$ and $\G^{pq}[X,\psi]$ are $r\times r$ matrices for which an
explicit expression is going to be determined below.

From~(\ref{eq_psians}) and~(\ref{eq_RkDef}) it follows that
\begin{eqnarray}
  \Psi & = & {\cal I} -\l \sum_{p,q=1}^m \mu_q \frac{T_p\, \G^{pq}\,
    \Tb_q}{\l-\mu_q}\, , \label{eq_psians2} \\
  \Psb & = & {\cal I} + \sum_{k,\ell=1}^m \frac{\ve\, T_\ell
    \ov{\G}^{k\ell} \Tb_k}{\l-\ve/\mb_\ell}\, . \label{eq_psibans}
\end{eqnarray}
Here we omitted the index $m$ in $\Psi_m$ and $\Psb_m$. In accordance
with~(\ref{eq_hermPsiPsidag}) we have to choose $\G^{pq}$ in such a form
that $\Psi\Psb$ will be independent of $\l$. A splitting into partial
fractions yields
\begin{equation}
  \label{eq_PsiPsibexp}
  \begin{split}
    \Psi\ov{\Psi} & = {\cal I} + \sum_{k,\ell} \frac{\ve T_\ell
      \ov{\G}^{k\ell} \Tb_k}{\l-\ve/\mb_\ell} - \l \sum_{p,q} \mu_q
      \frac{T_p\G^{pq}\Tb_q}{\l-\mu_q} - \sum_{p,q,k,\ell}
      (\l - \mu_q + \mu_q) \ve\mu_q \frac{T_p\G^{pq}\Tb_q\,
      T_\ell\ov{\G}^{k\ell}\Tb_k}{(\l-\mu_q)(\l-\ve/\mb_\ell)} \\
    & = {\cal I} + \sum_{k,\ell} \frac{\ve T_\ell \ov{\G}^{k\ell} \Tb_k}
      {\l-\ve/\mb_\ell} - \l \sum_{p,q} \mu_q \frac{T_p\G^{pq}\Tb_q}
      {\l-\mu_q} - \sum_{p,q,k,\ell} \frac{\ve\mu_q T_p\G^{pq}\Tb_q
      \,T_\ell\ov{\G}^{k\ell}\Tb_k}{\l-\ve/\mb_\ell} \\
    & \quad {} - \sum_{p,q,k,\ell} \frac{\ve \mu_q^2 \mb_\ell}
      {\mu_q \mb_\ell-\ve} \left( \frac{1}{\l-\mu_q} - \frac{1}
      {\l-\ve/\mb_\ell} \right) T_p\G^{pq}\Tb_q\,
      T_\ell\ov{\G}^{k\ell}\Tb_k \, .
  \end{split}
\end{equation}
This motivates us to define
\begin{equation}
  \Gat_{q\ell} := -\ve\mu_q\frac{\Tb_q T_\ell}{\mu_q\mb_\ell-\ve}\, ,
  \label{eq_GatDef}
\end{equation}
and, as the matrix $\G=(\G^{pq})$ has not yet been specified, to take
it to be inverse to $\Gat=(\Gat_{q\ell})$,
\begin{equation}
  \sum_{q=1}^m \G^{pq}\Gat_{q\ell} = \de^p{}_\ell {\cal I} \, .
    \label{eq_GDef}
\end{equation}
Upon insertion of eqs.~(\ref{eq_GatDef}) and~(\ref{eq_GDef}) into~%
(\ref{eq_PsiPsibexp}) nearly all terms cancel each other and we are left
with
\begin{equation}
  \Psi\ov{\Psi} = {\cal I} - \sum_{p,q} \mu_q T_p \G^{pq} \Tb_q =
    e^{-\P} \, . \label{eq_PsiPsibexp3}
\end{equation}
This expression is independent of~$\l$ and, therefore, we can identify
it with $e^{-\P}$ as in~(\ref{eq_hermPsiPsidag}). We see that for the
above choice of the $\G$-matrices the reality condition is satisfied,
and the solution $e^\P$ of eq.~(\ref{eq_BSSFTeom}) is parametrized by
the matrix-valued string fields $T_k$, $k=1,\ldots, m$. Note that~%
(\ref{eq_PsiPsibexp3}) coincides with $\Psi\ov{\Psi}|_{\l=\infty}=
\Psi|_{\l=\infty}$.

\noindent
{\bf Pole structure.} We are now going to exploit eq.~%
(\ref{eq_DrApprLin2}) in combination with the ansatz~(\ref{eq_psians2}).
First, it is easy to show that
\begin{equation}
  \Psi|_{\l=\frac{\ve}{\mb_k}} T_k = \left( {\cal I} + \sum_{p,q}
    \ve\mu_q \frac{T_p \G^{pq} \Tb_q}{\mu_q\mb_k-\ve} \right) T_k =
    T_k - \sum_{p,q} T_p\G^{pq}\Gat_{qk} = 0 \label{eq_PsiRes}
\end{equation}
and
\begin{equation}
  \ov{T_k}\, \ov{\Psi}|_{\l=\mu_k} = \Tb_k \left( {\cal I} + \sum_{p,q}
    \ve\frac{\mb_q T_q \ov{\G^{pq}}\, \Tb_p}{\mu_k\mb_q-\ve} \right)
    = \Tb_k -\sum_{p,q} \ov{\Gat}_{qk} \ov{\G}^{pq}\, \Tb_p = 0\, .
\end{equation}
Second, note that the right hand side of eq.~(\ref{eq_DrApprLin2})
is independent of $\l$ and therefore the poles on the left hand side have
to be removable. Putting to zero the corresponding residue at $\l=\frac{\ve}
{\mb_k}$ we obtain, due to~(\ref{eq_PsiRes}),
\begin{equation}
\Psi|_{\l=\frac{\ve}{\mb_k}} \left\{ (\ve\mb_k\Gtp + \Gp) T_k\right\} \sum_\ell
    \ov{\G}^{\ell k} \Tb_\ell = 0 \, .
\end{equation}
Obviously, a sufficient condition for a solution is
\begin{equation}
  \left( \Gtp + \frac{\ve}{\mb_k}\, \Gp\right) T_k = T_k {\cal Z}_k\, ,
    \label{eq_TkZk}
\end{equation}
with an arbitrary operator ${\cal Z}_k$ having the same Grassmann content as
the operator $\Gtp\big(\frac{\ve}{\mb_k}\big)$. In the same way,
the residue at~$\l=\mu_k$ should vanish, \\
\parbox{15cm}{
\begin{gather*}
  \left(\sum_p \mu_k T_p \G^{pk}\right) \left\{ \left( \frac{1}{\mu_k}
    \Gtp + G^+ \right) \Tb_k \right\} \Psb|_{\l=\mu_k} = 0 \\
  \Rightarrow \: (\Gtp + \mu_k\Gp)\Tb_k = {\cal Z}'_k \Tb_k \, ,
\end{gather*}}
\hfill
\parbox{8mm}{\begin{equation} \label{eq_TbkZpk} \end{equation}} \\
with another Grassmann-odd operator ${\cal Z}'_k$. Comparing
eqs.~(\ref{eq_TkZk}) and~(\ref{eq_TbkZpk}), we learn that
\begin{equation}
  \ov{{\cal Z}}_k \, \Tb_k = \ve {\cal Z}'_k \Tb_k \quad\Rightarrow\quad
    {\cal Z}'_k=\ve \ov{{\cal Z}}_k \, .
\end{equation}
In other words, eqs.~(\ref{eq_TbkZpk}) are not independent but follow
from eq.~(\ref{eq_TkZk}) by conjugation. For every collection $\{T_k,
k=1,\ldots,m\}$ of solutions to eqs.~(\ref{eq_TkZk}) we can determine
a solution to eq.~(\ref{eq_BSSFTeom}) from eqs.~(\ref{eq_GatDef})--%
(\ref{eq_PsiPsibexp3}).

\noindent
{\bf Projectors.} Now let us consider the simplest case $m=1$. Then,
eq.~(\ref{eq_psians2}) simplifies to
\begin{equation}
  \Psi = {\cal I} + \frac{\l\ve (|\mu|^2-\ve)}{\l-\mu} \, P \, ,
\end{equation}
where $P:=T(\ov{T}T)^{-1}\ov{T}$ is a hermitian projector, $P^2=P=\ov{P}$,
parametrized by an $n\times r$ matrix $T$. In the abelian ($n=1$) case $r$ is
the rank of the projector $P$ in the Hilbert space ${\cal H}$ of string
field theory. In the nonabelian ($n>1$) case $r\leq n$ can be identified
with the rank of the projector in the $u(n)$ factor of the $u(n)\otimes
{\cal H}$ Hilbert space.

From the extended solution $\Psi$ we obtain the solution
\begin{equation}
  e^{-\P} = \Psi|_{\l=\infty} = {\cal I} - (1-\ve|\mu|^2) P \label{eq_m1sol}
\end{equation}
of the equation of motion~(\ref{eq_BSSFTeom}). Thus, the simplest solutions
are parametrized by projectors in the string field theory Hilbert space.

To conclude this section, we summarize the main idea of the dressing approach
as follows: One has to extend the string field theory Hilbert space $u(n)
\otimes{\cal H}$ to $u(n)\otimes{\cal H}\otimes \C[\l,\l^{-1}]$, there
solve the equations on the extended string field $\P[X,\ps,\l]$ such that
$\Psi^{-1}[X,\ps,\l] = e^{\P[X,\ps,\l]}$, and then project back onto $u(n)
\otimes{\cal H}$. In this way, one obtains a solution $e^\P = \Psi^{-1}
(\l=\infty)$ of the initial equation of motion~(\ref{eq_BSSFTeom}), where the
extended solution $\Psi$ is parametrized by $T_k[X,\ps,\l=\ve\mb_k^{-1}]$
with $k=1,\ldots,m$.

\smallskip
\section{Solutions of the linear equations} \label{sec_sol}
\noindent
{\bf $\bsy{\Gtp(\l)}$-exact solutions.} In the previous section we have shown
that in the dressing approach solving the nonlinear string field equation~%
(\ref{eq_BSSFTeom}) reduces to solving the linear equations~(\ref{eq_TkZk}).
Solutions $T_k$, $k=1,\ldots,m$, of these equations parametrize solutions
$e^\P$ of eq.~(\ref{eq_BSSFTeom}) (cf.~(\ref{eq_PsiPsibexp3})). Obviously,
for obtaining some examples of solutions it is sufficient to find solutions
for ${\cal Z}_k=0$,
\begin{equation}
  \Gtp(\ve\mb_k^{-1}) T_k \equiv \left( \Gtp + \frac{\ve}{\mb_k}\,\Gp
    \right) T_k = 0\, . \label{eq_TkwoZk}
\end{equation}
Here, we present two classes of solutions to these equations.

Recall that $(\Gtp(\l))^2=0$ and, therefore,
\begin{equation}
  T_k = \Gtp(\ve\mb_k^{-1}) W_k \label{eq_Tkexact}
\end{equation}
is a solution of eq.~(\ref{eq_TkwoZk}) for any string field $W_k\in\,$Mat%
$(n\times r,\C)\otimes {\cal H}$. These solutions are in general nontrivial
because they are not annihilated by \Gp and \Gtp separately.

This discussion is valid for both \Ne1 strings ($\ve=-1$) and \N2
strings ($\ve=1$). Substituting~(\ref{eq_Tkexact}) into~(\ref{eq_PsiPsibexp3}),
we get explicit solutions $e^\P$. In the \Ne1 case, other obvious solutions are
all BRST-closed vertex operators $T_k$ in the small Hilbert space of~%
\cite{Friedan:1985ge} as they satisfy $[Q,T_k]=0$ and $[\eta_0, T_k]=0$
separately.

For the case of \N2 strings we will discuss two classes of explicit
solutions of~(\ref{eq_TkZk}) (for both ${\cal Z}_k=0$ and ${\cal Z}_k\neq 0$)
which in general do not have the form~(\ref{eq_Tkexact}).

\noindent
{\bf $\bsy{N\,{=}\,2}$ string solutions for $\bsy{{\cal Z}_k=0}$.} We already
presented the realization of \Gp and \Gtp in terms of the constituents of an \N2
matter multiplet in eqs.~(\ref{eq_N4SCA_N2_1}) and~(\ref{eq_N4SCA_N2_2}). Using
this realization, one can factorize the ``$G$-matrix'' in~(\ref{eq_G_matrix})
according to
\begin{equation}
  \begin{pmatrix} \Gp & \Gtp \\ 
  \Gtm & \Gm \end{pmatrix} =
  \begin{pmatrix} \ps^{+1} & -\ps^{+0} \\ 
  -\ps^{-\nb} & \ps^{-\ob} \end{pmatrix}
  \begin{pmatrix} \pa \Zb^\ob & \pa Z^0 \\ 
  \pa \Zb^\nb & \pa Z^1 \end{pmatrix} \, .
  \label{eq_G_fact}
\end{equation}
As in section~\ref{sec_RealProp}, this matrix transforms under current
$SU(1,1)$-rotations acting from the left (note that the world-sheet fermions are
charged under the current group) and under the additional $SU(1,1)$-rotations 
as in~(\ref{eq_GuDef}) acting from the right. 
The latter transform~(\ref{eq_G_fact}) to
\begin{equation}
  \begin{pmatrix} \Gp(u) & \Gtp(u) \\ 
  \Gtm(u) & \Gm(u) \end{pmatrix} =
  \begin{pmatrix} \ps^{+1} & -\ps^{+0} \\ 
  -\ps^{-\nb} & \ps^{-\ob} \end{pmatrix}
  \begin{pmatrix} \pa \Zb^\ob & \pa Z^0 \\ 
  \pa \Zb^\nb & \pa Z^1 \end{pmatrix}
  \begin{pmatrix} u_1 & \ub_2 \\ 
  u_2 & \ub_1 \end{pmatrix} \, .
\end{equation}
By right-multiplication with $\left( \begin{smallmatrix} u_1^{-1}
& 0 \\ 0 & \ub_1^{-1} \end{smallmatrix} \right)$ as in~(\ref{eq_GlDef})
we can express everything in terms of $\l$,
\begin{equation}
  \begin{pmatrix} \Gp(\l) & \Gtp(\l) \\ 
  \Gtm(\l) & \Gm(\l) \end{pmatrix} =
  \begin{pmatrix} \ps^{+1} & -\ps^{+0} \\ 
  -\ps^{-\nb} & \ps^{-\ob} \end{pmatrix}
  \begin{pmatrix} \pa \Zb^\ob(\l) & \pa Z^0(\l) \\ 
  \pa \Zb^\nb(\l) & \pa Z^1(\l)
  \end{pmatrix} \, ,
\end{equation}
where the coordinates
\begin{equation}
  Z^0(z,\zb,\l) := Z^0(z,\zb) + \l\Zb^\ob(z,\zb) \quad\mbox{and}\quad
  Z^1(z,\zb,\l) := Z^1(z,\zb) + \l\Zb^\nb(z,\zb) \label{eq_ZDef}
\end{equation}
define a new complex structure on the target space $\C^{1,1}$~%
\cite{Berkovits:1994vy, Lechtenfeld:1999ik}. From eq.~(\ref{eq_ZZbOPE})
we derive that $Z^a(z,\zb,\l)$ are null coordinates:
\begin{equation}
  Z^a(z,\zb,\l) Z^b(w,\wb,\l) \sim 0 \, . \label{eq_ZlZlOPE}
\end{equation}
The derivation of the string field algebra in~(\ref{eq_TkZk}) can be written
entirely in terms of $Z^a(z,\zb,\l)$,
\begin{equation}
  \Gtp(z,\l) = \Gtp(z) + \l\Gp(z) = -\ve_{ab} \ps^{+a}(z)\, \pa
    Z^b(z,\l)\, ,
\end{equation}
and from~(\ref{eq_ZlZlOPE}) it follows that
\begin{equation}
  \oint \frac{dw}{2\pi i} \Gtp(w,\l)
    Z^c(z,\zb,\l) = 0
\end{equation}
with the integration contour running around $z$. This equation implies
that every analytic functional $T_k$ of the new spacetime coordinates $Z^0
(z,\zb,\mb_k^{-1})$ and $Z^1(z,\zb,\mb_k^{-1})$ solves~(\ref{eq_TkwoZk}).
Indeed, it can be easily checked that for any integer $p,q$, we have
\begin{equation}
  \oint \frac{dw}{2\pi i} \Gtp\left(w,\frac{1}{\mb_k}\right)
    :(Z^0)^p (Z^1)^q(z,\zb,\mb_k^{-1}) :\; = 0 \, .
\end{equation}
The functional $T_k$ may also depend on arbitrary derivatives $\pa^\ell
Z^a(z,\mb_k^{-1})$ (note that, for $\ell=1$, $\pa Z^a(z,\mb_k^{-1})$ is
$\Gtp(\mb_k^{-1})$-exact). Due to~(\ref{eq_ZZbOPE}), it may furthermore depend
on $\ps^{+a}(z)$ or its derivatives. Given some analytic functionals 
$T_k [Z^a(z,\zb,\mb_k^{-1}),\pa^\ell Z^a(z,\mb_k^{-1}),\ps^{+a}(z),
\pa^p \ps^{+a}(z)]$
with values in Mat$(n\times r,\C)$ for $k=1,\ldots,m$, we can determine a
solution of~(\ref{eq_BSSFTeom}) with the help of eq.~(\ref{eq_PsiPsibexp3}).
Note that we do not claim to have found all solutions.

\noindent
{\bf $\bsy{N\,{=}\,2}$ string solutions for $\bsy{{\cal Z}_k\neq 0}$.} 
We restrict
ourselves to the abelian case $n=1$. In addition to the coordinates
$Z^a(z,\zb,\mb_k^{-1})$ from above, we introduce vertex operators
\begin{equation}
  Y^0(z,\zb,\mb_k^{-1}) := \tfrac{1}{2} \big( \Zb^\ob(z,\zb) - \tfrac{1}{\mb_k}
     Z^0(z,\zb) \big) \quad\mbox{and}\quad Y^1(z,\zb,\mb_k^{-1}) :=
     \tfrac{1}{2} \big( \Zb^\nb(z,\zb) - \tfrac{1}{\mb_k} Z^1(z,\zb) \big) .
\end{equation}
They satisfy the following OPE with $Z^a(z,\zb,\mb_k^{-1})$:
\begin{equation}
  Z^a(z,\zb,\mb_k^{-1})\, Y^b(w,\wb,\mb_k^{-1}) \sim -2\ve^{ab}\ln |z-w |^2 
  \, .
\end{equation}
From this, we immediately derive
\begin{equation}
   \oint \frac{dw}{2\pi i} \Gtp\left(w,\frac{1}{\mb_k}\right)
    : e^{\a_a^k Y^a} (z,\zb,\mb_k^{-1}) : \;
   = -2 : e^{\a_a^k Y^a} (z,\zb,\mb_k^{-1}) : \,\a^k_b \ps^{+b}(z) \, ,
\end{equation}
where $\a_a^k$ are complex constants. We see that for $k=1,\ldots,m$,
\begin{equation}
  T_k = \; : e^{\a_a^k Y^a} (z,\zb,\mb_k^{-1}) :
\end{equation}
satisfy eq.~(\ref{eq_TkZk}) and therefore 
produce a solution of~(\ref{eq_BSSFTeom}) via~(\ref{eq_PsiPsibexp3}).

\smallskip
\section{Conclusions} \label{sec_Concl}
\noindent
In this paper we have demonstrated that the equation of motion for Berkovits'
WZW-like string field theory is integrable. Two approaches to generating
solutions of this equation were discussed and adapted to string field
theory: the splitting technique and the dressing method. In essence, both
procedures reduce the nonpolynomial equation of motion to some linear
equations. The solutions of these linear equations give us nonperturbative
solutions of the original equation of motion. Our discussion was kept
general enough to apply to the case of \Ne1 superstrings as well as to the
case of \N2 strings.

In order to demonstrate the power of our methods we explicitly constructed
some solutions to the linear equations via the dressing approach. For \Ne1
superstrings, a quite general class of solutions was presented; for \N2
strings, the same and additional classes of solutions were found. Following
the recipe given in section~\ref{sec_DressAppr}, one can easily translate all
these to classical configurations of Berkovits' (super)string field theory.

A lot of work remains to be done: In order to establish which among our
solutions represent soliton-like objects within the theory, one has to
evaluate their energy. It would be interesting to find criteria on the~$T_k$
for the solution to be a soliton, an instanton, or a monopole.
In the case of \N2 strings explicit solitonic solutions to the
corresponding field theory equations have been constructed earlier~%
\cite{Lechtenfeld:2001aw, Lechtenfeld:2001uq, Lechtenfeld:2001gf}; it
is plausible that they can be promoted to the string level. An examination
of the fluctuations around these nonperturbative solutions should determine
what kind of object they represent in string theory. If some of these
solutions turn out to describe D-branes, perhaps another check of Sen's
conjecture on the relation between the tension of D-branes and the string
field theory action is feasible. Due to our choosing the simplest ans\"atze
for the splitting and the dressing methods, we have obtained not the
broadest classes of field configurations. However, nothing prevents one
from employing more general ans\"atze and thereby creating more general
solutions for \Ne1 superstring field theory.

It is also desirable to elucidate the geometrical interpretation of all
these classical configurations. To this end, we hope that further work
will shed some light on the physical meaning of the projector $P$, which was
central to the ansatz for the simplest solution to the linear equations
(cf.\ section~\ref{sec_DressAppr}). Another direction for future
investigation could be the transfer of our analysis to cubic superstring
field theory. The knowledge of nonpertubative solutions should help to
understand the relation of Berkovits' nonpolynomial with Witten's cubic
superstring field theory. In all instances, it would be exciting to make
contact with the current discussion of supersliver states.

\bigskip
\noindent
{\large{\bf Acknowledgements}}

\smallskip\noindent
We acknowledge discussions with M.~Olshanetsky. S.~U.\ would like to thank
N.~Berkovits for useful discussions. This work is partially supported by
DFG grant Le~838/7-1 and by a sabbatical research grant of the Volkswagen-%
Stiftung.


\bigskip

\begin{thebibliography}{99}

\bibitem{Witten:1985cc}
E.~Witten, {\it Noncommutative geometry and string field theory},
Nucl.\ Phys.\ B~{\bf 268}~(1986)~253.

\bibitem{Witten:1986qs}
E.~Witten, {\it Interacting field theory of open superstrings},
Nucl.\ Phys.\ B~{\bf 276}~(1986)~291.

\bibitem{Preitschopf:fc}
C.~R.~Preitschopf, C.~B.~Thorn and S.~Yost, {\it Superstring field theory},\\
Nucl.\ Phys.\ B~{\bf 337}~(1990)~363.

\bibitem{Arefeva:1989cp}
I.~Ya.~Aref'eva, P.~B.~Medvedev and A.~P.~Zubarev, {\it New representation
for string field solves the consistence problem for open superstring field},
Nucl.\ Phys.\ B~{\bf 341}~(1990)~464.

\bibitem{Thorn:1988hm}
C.~B.~Thorn, {\it String field theory},
Phys.\ Rept.\ {\bf 175}~(1989)~1.

\bibitem{Friedan:1985ge}
D.~Friedan, E.~Martinec and S.~Shenker, 
{\it Conformal invariance, supersymmetry and string theory},
Nucl.\ Phys.\ B~{\bf 271}~(1986)~93.

\bibitem{Wendt:1987zh}
C.~Wendt, {\it Scattering amplitudes and contact interactions in Witten's
superstring field theory},
Nucl.\ Phys.\ B~{\bf 314}~(1989)~209.

\bibitem{Lechtenfeld:1988tr}
O.~Lechtenfeld and S.~Samuel, {\it Gauge invariant modification of Witten's
open superstring},\\
Phys.\ Lett.\ B~{\bf 213}~(1988)~431.

\bibitem{Berkovits:1991gj}
N.~Berkovits, M.~Hatsuda and W.~Siegel, {\it The big picture},
Nucl.\ Phys.\ B~{\bf 371}~(1992)~434 [hep-th/9108021].

\bibitem{Berkovits:1995ab}
N.~Berkovits, {\it Super-Poincar\'{e} invariant superstring field theory},
Nucl.\ Phys.\ B~{\bf 450}~(1995)~90, Erratum ibid.\ B~{\bf 459}~(1996)~439
[hep-th/9503099].

\bibitem{Ohmori:2001am}
K.~Ohmori, 
{\it A review on tachyon condensation in open string field theories},
hep-th/0102085.

\bibitem{DeSmet:2001af}
P.-J.~De~Smet, {\it Tachyon condensation: Calculations in string field theory},
hep-th/0109182.

\bibitem{Sen:1999mg}
A.~Sen, {\it Non-BPS states and branes in string theory},
hep-th/9904207.

\bibitem{Berkovits:2000zj}
N.~Berkovits, {\it The tachyon potential in open Neveu-Schwarz string field
theory},\\
JHEP {\bf 0004}~(2000)~022 [hep-th/0001084].

\bibitem{Berkovits:2000hf}
N.~Berkovits, A.~Sen and B.~Zwiebach, {\it Tachyon condensation in superstring
field theory},\\
Nucl.\ Phys.\ B~{\bf 587}~(2000)~147 [hep-th/0002211].

\bibitem{DeSmet:2000dp}
P.~J.~De Smet and J.~Raeymaekers, {\it Level four approximation to the tachyon
potential in superstring field theory},
JHEP {\bf 0005}~(2000)~051 [hep-th/0003220].

\bibitem{Iqbal:2000st}
A.~Iqbal and A.~Naqvi, {\it Tachyon condensation on a non-BPS D-brane},
hep-th/0004015.

\bibitem{Ohmori:2001sx}
K.~Ohmori, {\it Tachyonic kink and lump-like solutions in superstring
field theory},\\
JHEP {\bf 0105}~(2001)~035 [hep-th/0104230].

\bibitem{Kluson:2001kk}
J.~Kluso\v{n}, {\it Proposal for background independent Berkovits' superstring
field theory},\\
JHEP {\bf 0107}~(2001)~039 [hep-th/0106107].

\bibitem{Sakaguchi:2001kk}
M.~Sakaguchi, {\it Pregeometrical formulation of Berkovits' open RNS
superstring field theories},
hep-th/0112135.

\bibitem{Marino:2001ny}
M.~Marino and R.~Schiappa, {\it Towards vacuum superstring field theory:
The supersliver},\\
hep-th/011223.

\bibitem{Arefeva:2002mb}
I.~Y.~Aref'eva, D.~M.~Belov and A.~A.~Giryavets, {\it Construction of the
vacuum string field theory on a non-BPS brane},
hep-th/0201197.

\bibitem{Ohmori:2002}
K.~Ohmori, {\it Comments on solutions of vacuum superstring field theory},
hep-th/0204138.

\bibitem{Kluson:2001sb}
J.~Kluso\v{n}, {\it Some remarks about Berkovits' superstring field theory},
JHEP {\bf 0106}~(2001)~045 [hep-th/0105319].

\bibitem{Kluson:2002kk}
J.~Kluso\v{n}, {\it Some solutions of Berkovits' superstring field theory},
hep-th/0201054.

\bibitem{Berkovits:1994vy}
N.~Berkovits and C.~Vafa, {\it \Ng4 topological strings},
Nucl.\ Phys.\ B~{\bf 433}~(1995)~123 \\ {}
[hep-th/9407190].

\bibitem{Berkovits:1993xq}
N.~Berkovits and C.~Vafa, {\it On the uniqueness of string theory},
Mod.\ Phys.\ Lett.\ A~{\bf 9}~(1994)~653 [hep-th/9310170].

\bibitem{Berkovits:1999im}
N.~Berkovits, C.~Vafa and E.~Witten, {\it Conformal field theory of AdS
background with Ramond-Ramond flux},
JHEP {\bf 9903}~(1999)~018 [hep-th/9902098].

\bibitem{Junemann:1999hi}
K.~J\"unemann, O.~Lechtenfeld and A.~D.~Popov, {\it Non-local symmetries
of the closed \N2 string},
Nucl.\ Phys.\ B~{\bf 548}~(1999)~449 [hep-th/9901164].

\bibitem{Bellucci:2001qi}
S.~Bellucci and A.~Galajinsky, {\it Can one restore Lorentz invariance in
quantum \N2 string?},
hep-th/0112024.

\bibitem{Arefeva:2001ps}
I.~Y.~Aref'eva, D.~M.~Belov, A.~A.~Giryavets, A.~S.~Koshelev and 
P.~B.~Medvedev,
{\it Noncommutative field theories and (super)string field theories},
hep-th/0111208.

\bibitem{Berkovits:1997pq}
N.~Berkovits and W.~Siegel, 
{\it Covariant field theory for self-dual strings},\\
Nucl.\ Phys.\ B~{\bf 505}~(1997)~139 [hep-th/9703154].

\bibitem{Lechtenfeld:2000qj}
O.~Lechtenfeld and A.~D.~Popov, {\it On the integrability of covariant field
theory for open \N2 strings},
Phys.\ Lett.\ B~{\bf 494}~(2000)~148 [hep-th/0009144].

\bibitem{Humphreys}
J.~E.~Humphreys, {\it Introduction to Lie algebras and representation theory},
Springer (1972).

\bibitem{Helgason}
S.~Helgason, {\it Differential geometry, Lie groups and symmetric spaces},
Academic Press (1978).

\bibitem{D'Adda:1987rx}
A.~D'Adda and F.~Lizzi, {\it Space dimensions from supersymmetry for the \N2
spinning string: A four-dimensional model},
Phys.\ Lett.\ B~{\bf 191} (1987) 85.

\bibitem{Atiyah:pw}
M.~F.~Atiyah and R.~S.~Ward, {\it Instantons and algebraic geometry},\\
Commun.\ Math.\ Phys.\ {\bf 55}~(1977)~177.

\bibitem{Lechtenfeld:2001ie}
O.~Lechtenfeld and A.~D.~Popov, {\it Noncommutative 't Hooft instantons},
JHEP {\bf 0203}~(2002)~040 [hep-th/0109209].

\bibitem{Nekrasov:1998ss}
N.~Nekrasov and A.~Schwarz, {\it Instantons on noncommutative $\R^4$ and
(2,0) superconformal six-dimensional theory},
Commun.\ Math.\ Phys.\ {\bf 198}~(1998)~689 [hep-th/9802068].

\bibitem{Berkovits:2001nr}
N.~Berkovits, {\it Review of open superstring field theory},
hep-th/0105230.

\bibitem{Zakharov:1979}
V.~E.~Zakharov and A.~B.~Shabat, {\it Integration of nonlinear equations
of mathematical physics by the method of inverse scattering.II},
Funct.\ Anal.\ Appl.\ {\bf 13}~(1979)~166.

\bibitem{Forgacs:1983gr}
P.~Forg\'{a}cs, Z.~Horv\'{a}th and L.~Palla, 
{\it Solution-generating technique for self-dual monopoles},
Nucl.\ Phys.\ B~{\bf 229}~(1983)~77.

\bibitem{Ward:vc}
R.~S.~Ward, {\it Classical solutions of the chiral model, unitons, and
holomorphic vector bundles},
Commun.\ Math.\ Phys.\ {\bf 128}~(1990)~319.

\bibitem{Lechtenfeld:2001aw}
O.~Lechtenfeld and A.~D.~Popov, {\it Noncommutative multi-solitons in 2+1
dimensions},\\
JHEP {\bf 0111}~(2001)~040 [hep-th/0106213].

\bibitem{Lechtenfeld:1999ik}
O.~Lechtenfeld and A.~D.~Popov, 
{\it Closed \N2 strings: Picture-changing, 
hidden symmetries and SDG hierarchy},
Int.\ J.\ Mod.\ Phys.\ A~{\bf 15}~(2000)~4191 [hep-th/9912154].

\bibitem{Lechtenfeld:2001uq}
O.~Lechtenfeld, A.~D.~Popov and B.~Spendig, 
{\it Noncommutative solitons in open \N2 string theory},
JHEP {\bf 0106}~(2001)~011 [hep-th/0103196].

\bibitem{Lechtenfeld:2001gf}
O.~Lechtenfeld and A.~D.~Popov, 
{\it Scattering of noncommutative solitons in 2+1 dimensions},
Phys.\ Lett.\ B~{\bf 523}~(2001)~178 [hep-th/0108118].

\end{thebibliography}
\end{document}